# Broadband Acoustic Cloak for Ultrasound Waves

Shu Zhang, Chunguang Xia and Nicholas Fang [1]
Department of Mechanical Science & Engineering and the Beckman Institute of Advanced Science and Technology, University of Illinois at Urbana-Champaign, Illinois 61801

Invisibility devices based on coordinate transformation have opened up a new field of considerable interest. Such a device is proposed to render the hidden object undetectable under the flow of light or sound, by guiding and controlling the wave path through an engineered space surrounding the object. We present here the first practical realization of a low-loss and broadband acoustic cloak for underwater ultrasound. This metamaterial cloak is constructed with a network of acoustic circuit elements, namely serial inductors and shunt capacitors. Our experiment clearly shows that the acoustic cloak can effectively bend the ultrasound waves around the hidden object, with reduced scattering and shadow. Due to the non-resonant nature of the building elements, this low loss (~6dB/m) cylindrical cloak exhibits excellent invisibility over a broad frequency range from 52 to 64 kHz in the measurements. The low visibility of the cloaked object for underwater ultrasound shed a light on the fundamental understanding of manipulation, storage and control of acoustic waves. Furthermore, our experimental study indicates that this design approach should be scalable to different acoustic frequencies and offers the possibility for a variety of devices based on coordinate transformation.

---

[1] nicfang@illinois.edu

Recently, a new design paradigm called coordinate transformation has inspired a series of key explorations to manipulate, store and control the flow of energy, in form of either sound, elastic waves or light radiation. In electromagnetism, because of the coordinate invariance of Maxwell's equations, the space for light can be bent in almost arbitrary ways by providing a desired spatial distribution of electric permittivity ε and magnetic permeability μ. (*1, 2*) A set of novel optical devices were proposed based on transformation optics (*3, 4, 5*); they usually call for complicated medium with anisotropic and spatially varying permittivity and permeability tensor to accomplish the desired functionality. Recent advances in synthetic structured metamaterial (*6, 7, 8*) whose properties are determined by its subwavelength structure, offers the potential to physically implement these complicated media. By modifying the shape and arrangement of these subwavelength constituent elements, anisotropy and spatial variation can be achieved in the artificial metamaterials.

Among the most exciting examples is perhaps an electromagnetic cloak that can render the objects invisible. The first experimental demonstration of such a cloak was reported in microwave using structured metamaterial composed of metallic resonant rings (9). However, the invisibility effect was only obtained in a narrow frequency range because of the strong dispersion inherent to the resonant elements used to build the cloak. In addition, such resonances led to undesired material absorption in the cloak. To mitigate these constrains, several different schemes of cloaking utilizing non-resonant structure were proposed (*10, 11, 12, 13, 14*).One example is a so-called carpet cloak which compresses the cloaked space into a thin sheet (*15, 16*). However, the waves travel faster through the carpet cloak than through the outer space, such a faster-than-light speed thereby sets a fundamental restriction for broadband application in ambient air. Therefore, the experiments of these carpet cloaks were so far conducted in a dielectric medium with higher index.

In contrast, cloaking of other classical waves such as acoustic waves do not suffer from such limitation for electromagnetic cloaks (*17, 18, 19*). However, in general the elastodynamic equations do not have this invariance symmetry as proven by Milton et al (*20*). Fortunately, acoustic waves in fluids follow such form invariance and several theoretical schemes of transformation have been proposed (*21, 22, 23, 24*). Theoretical analysis of an acoustic cloak (*24*) was reported based on the equivalence between transverse electric electromagnetic waves and acoustic waves in a two-dimensional (2D) geometry. Yet, this 2D acoustic cloak requires anisotropic mass density which is not common in naturally-occurring materials (*25, 26*). Consequently the experimental studies of acoustic cloak have been hampered by the difficulty in creating suitable materials and so far remain challenging.

In this paper, we overcome the above challenges in acoustic cloak design by introducing an acoustic transmission line approach. By taking the analogy between lumped acoustic elements and electronic circuit elements, this transmission line (TL) approach enabled a new class of acoustic metamaterials (*27*) and ultrasound focusing through a metamaterial network (*28*). As a demonstration, we designed a 2D cylindrical cloak as shown in Fig.1 in order to hide an object in the center. This acoustic cylindrical cloak is implemented by a 2D array of sub-wavelength cavities and connecting channels with spatially tailored geometry. The acoustic wave propagation through this discrete network can be described by a set of telegrapher's equations in which the motion of the fluid is equivalent to the behavior of the current in the circuit. This approach enables the realization of acoustic metamaterials with simple structure, ease of manufacturing and scaling, offering the potential to achieve a variety of acoustic devices based on transformation. Moreover, the acoustic cloak is expected to be low-loss and broadband with the use of non-resonant constituent elements.

In our design, the 2D acoustic metamaterial cloak is designed to squeeze the cylindrical region $0<r<R_2$ into an annular region $R_1<r'<R_2$ where r and r' are the radial coordinate in the original and transformed system respectively. The acoustic waves are thus excluded from the extended volume and smoothly bent inside the cloak, with no perturbation of exterior field. We choose the inner and outer radius of the cloak as $R_1$=13.5mm, $R_2$=54.1mm. Using 2D telegrapher's equations to describe the distributed acoustic system, this given warping of space can be achieved by providing the desired distribution of the serial inductor and shunt capacitor in the annular region interpreted as $L_r = \rho_w \frac{\Delta r}{2S_r}$  $L_\phi = \rho_w \frac{r\Delta\phi}{2S_\phi}\left(\frac{r-R_1}{r}\right)^2$  $C = 2\Delta r S_\phi \beta_w \left(\frac{R_2}{R_2-R_1}\right)^2$ as plotted in Fig.2(b). To facilitate the experimental realization of the cloak, here we used the simplified functional form of the cloaking parameters.

The acoustic cloak with the above parameter specification is physically synthesized by a planar network of acoustic circuits machined in an aluminum plate as shown in Fig.2 (a). These building blocks are cascaded in a lattice configuration which is diagonal in a cylindrical basis. In such a topology, the cloak is approximated by sixteen homogeneous concentric cylinders. From the first to the fourth layers, the unit cell size along radial direction is λ/7, λ/8, λ/9 and λ/9 (λ is the wavelength in water at 60 kHz) respectively. Afterwards, the cylinder layers are evenly spaced with the distance equal to λ/10 along the radial direction. On the other hand, the first cylinder near the inner lining of the cloak is divided to 32 units around the circumference. To keep the size of the constituent element smaller than λ/10 along circumferential direction, starting from the second layer, the number of cells is doubled to 64, and further increased to 128 from the sixth layer.

In our experiment, the water-filled network structure behaves as a lumped anisotropic TL for incoming underwater ultrasound. In each unit cell as shown in Fig.2 (b), as the size of each unit cell is only around one-tenth of the wavelength at operating frequency of 60 kHz, the cavity with large volume in center works as an acoustic capacitor $C = \frac{V}{\rho_w c_w^2}$ whereas the channels connecting it to the four neighboring cavities act as serial inductors $L_r = \rho_w \frac{l_r}{S_r}$ $L_\phi = \rho_w \frac{l_\phi}{S_\phi}$ (*29,30,31*). These relations imply the realization of the TL cloak with the spatially varying parameter profile by tailoring the geometry of the corresponding building blocks as listed in Fig.2(c). Such an anisotropic circuit network reroutes the paths of underwater sound around the cloaked object without significant scattering. In the lumped circuit model, the aluminum is assumed as acoustically rigid, which is a good approximation considering the acoustic impedance $\rho c$ of aluminum is around eleven times of that of water. A more careful analysis including elasticity of the solid suggested that at low frequency the majority of acoustic energy can be predominantly confined in the fluid, when such an excitation originates from the liquid. (*32*) On the other hand, aluminum does participate in the wave propagation, and may increases the loss of cloaking material (*33*).

For experimental confirmation of the cloaking performance, we placed an object in a water tank and compared the wavefronts of propagating ultrasound in our measurement, with or without the presence of our cloak as shown in Fig.1. The object is a steel cylinder with size equal to the inner radius of the cloak. The side of the cloak machined with the metamaterial network was placed against the bottom of the tank in order to seal water inside. The ultrasound waves were launched from a spherical shaped transducer as a point source with distance of 165mm (about 6.5 wavelengths) away from the center of the cloak. To map the pressure field, a hydrophone was mounted on a horizontal linear translation stages to scan in x-y directions. By stepping the hydrophone in small increments of 3mm and recording the acoustic pulse signal from the water at every step, we acquired the 2D spatial field distribution of the ultrasound wave scattering pattern.

It is evident in our experiment that the presence of steel cylinder alone in the water tank produces considerable scattering and shadowing at 60 kHz as shown in Fig. 3(a). By surrounding the steel cylinder with the metamaterial cloak in Fig. 3(d), however, the wave trajectory was restored behind the cloak with diminutive distortion in the cylindrical wavefronts, making the cloak and the hidden cylinder invisible under the hydrophone. Very small attenuation of the transmitted fields is observed on the exit side of the cloak, demonstrating the low-loss nature of the metamaterial cloak based on transmission line model.

To demonstrate the broadband nature of our designed cloak, the acoustic wave field distributions at 52 kHz and 64 kHz are presented in Fig. 3 (b) (c) (e) (f) for both cases with and without cloak. The field maps from these measurements present similar cloaking behavior with those at 60 kHz. This is not surprising since our metamaterial cloak is constructed by non-resonant elements. Theoretically, the cloak is expected to operate over a wide frequency range of 40 to 80 kHz. In fact, at frequency below 40 kHz, the scattering from the object with radius of 13.5mm become negligible. At the same time, the effectiveness of cloak at high frequency is restricted by two factors. The first is the breakdown of the lumped circuit model approximation at 120 kHz when the unit cell is comparable to one quarter of wavelength. By using smaller size of unit cells, this limit can be lifted to higher frequency. The other limit is the cutoff frequency at 80 kHz due to the low-pass topology of the circuit network. By modifying the geometry of the building block, this cutoff frequency can be potentially extended. However, in the current experiment we can only verify the cloaking behavior from 52 to 64 kHz in the experiment due to the limited operating frequency range of the transducer.

To further quantify the reduction of scattering and shadowing of the cloaked object, we conducted a set of measurements over different frequencies. The peak values of pressure along the wavefronts behind the cloak were obtained using a MATLAB program to process the experimental data. To facilitate the comparison of the cloaking performance, we defined the averaged visibility of an object as $\bar{\gamma} = \frac{1}{n}\sum_{j=1}^{n}\gamma_j$ , where $\gamma_j = \frac{P_{max,j} - P_{min,j}}{P_{max,j} + P_{min,j}}$ , $P_{max,j}$ and $P_{min,j}$ are the maximum and minimum peak values along the wavefront numbered by $j$. This can be compared to the traditional measurement of so-called scattering cross-sectional area, but performed for the convenience of the near field measurement and limited field of view in our experiment setup. Fig. 4(a) shows one example of the measured peak pressure at 60 kHz along one wavefront on the exit side of the object for both cases with and without cloak. This wavefront is near the boundary of the cloak between $y = 100mm$ and $y = 170mm$ .As reference, the measurement results for the free space when there is neither object nor cloak in the water tank is plotted. There is small modulation in the amplitude along one wavefront. In Fig.4 (b), the averaged visibility of the cloaked object over all the wavefronts on the exit side is compared with the one with only bare cylinder. The comparison clearly indicates that the cloak preserves good shielding effectiveness over a broad frequency range even with impedance mismatch at the outer interface of the cloak. We can read the visibility of 0.62 for the bare steel cylinder, whereas the visibility of the cloaked steel cylinder is reduced to 0.32 at 60 kHz, showing significant reduction in scattering and shadowing. Ideally, the averaged visibility should be zero for the case when there is no scatter However, due to the noise in the measurement, the averaged visibility for the free space still has a small value, as shown in Fig.4 (b).

In conclusion, we have demonstrated a 2D acoustic cloak that can significantly reduce the visibility of the hidden object from underwater acoustic waves. This underwater acoustic cloak can be readily implemented by a network of anisotropic acoustic transmission line. Such a new class of acoustic metamaterial is built from non-resonant circuit elements and can work over a broad frequency range. Moreover, this transmission line approach may have potential applications for a myriad of fascinating devices beyond cloaking based on coordinate transformation.

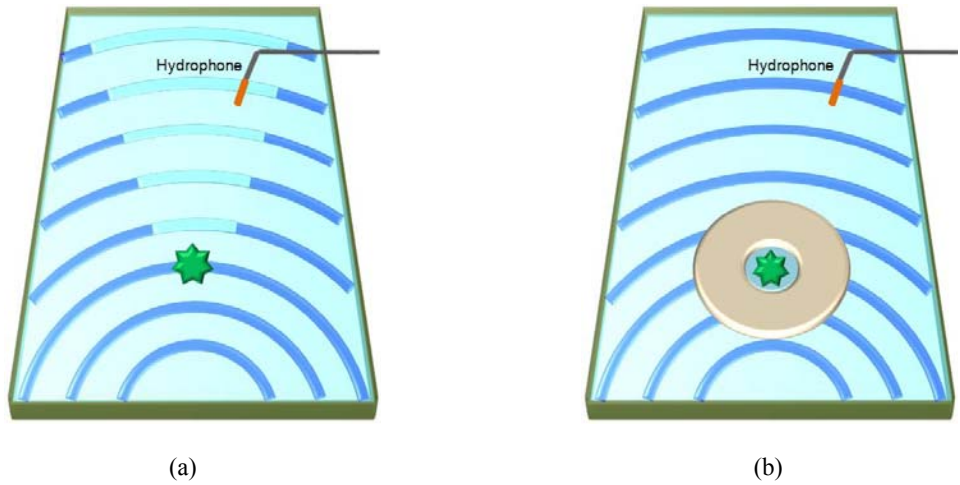

(a)                                          (b)

**Fig. 1.** Schematic diagram of the experimental setup. A burst of monotonic signal with a width of twenty periods was used to drive the transducer as an underwater point source in the water tank. One needle-sized hydrophone detected the ultrasonic signals in the immediate environment of (a) the object and (b) the cloaked object.

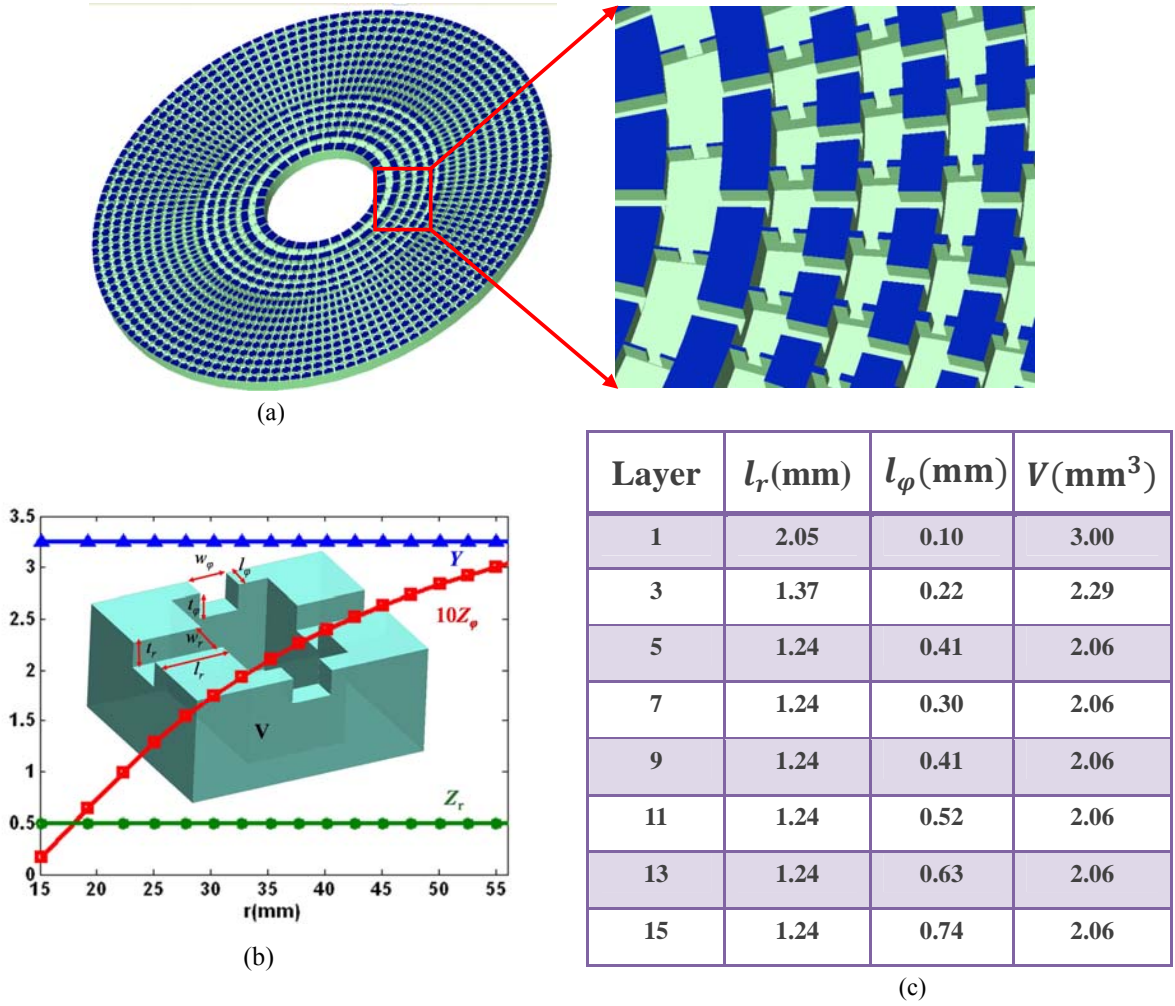

| Layer | $l_r$(mm) | $l_\varphi$(mm) | $V$(mm³) |
|---|---|---|---|
| 1 | 2.05 | 0.10 | 3.00 |
| 3 | 1.37 | 0.22 | 2.29 |
| 5 | 1.24 | 0.41 | 2.06 |
| 7 | 1.24 | 0.30 | 2.06 |
| 9 | 1.24 | 0.41 | 2.06 |
| 11 | 1.24 | 0.52 | 2.06 |
| 13 | 1.24 | 0.63 | 2.06 |
| 15 | 1.24 | 0.74 | 2.06 |

(c)

**Fig. 2**. A 2D acoustic cloak for underwater ultrasound waves. (a) The configuration of the acoustic cylindrical cloak synthesized by an acoustic transmission line, namely serial inductors and shunt capacitors. The inset is the expanded view of the network. The cavities with large volume work as shunt capacitors and those cavities are connected by narrow channels that act as the serial inductors. (b) One building block of the acoustic circuit, each unit cell consists of one large cavity in the center with channels connecting to the four neighboring blocks. The reduced cloaking parameters are used in the design. The serial impedance $Z_r$ , shunt admittance Y have constant values and $Z_\varphi$ increases as radius changes from $R_1$=13.5mm to the $R_2$=54.1mm. (c) The geometry parameters of the building blocks in the layers with odd number are presented in the table. The depth and width $t_r, w_r$ and $t_\varphi, w_\varphi$ of the channels along radial and angular directions have constant values of 0.5mm.

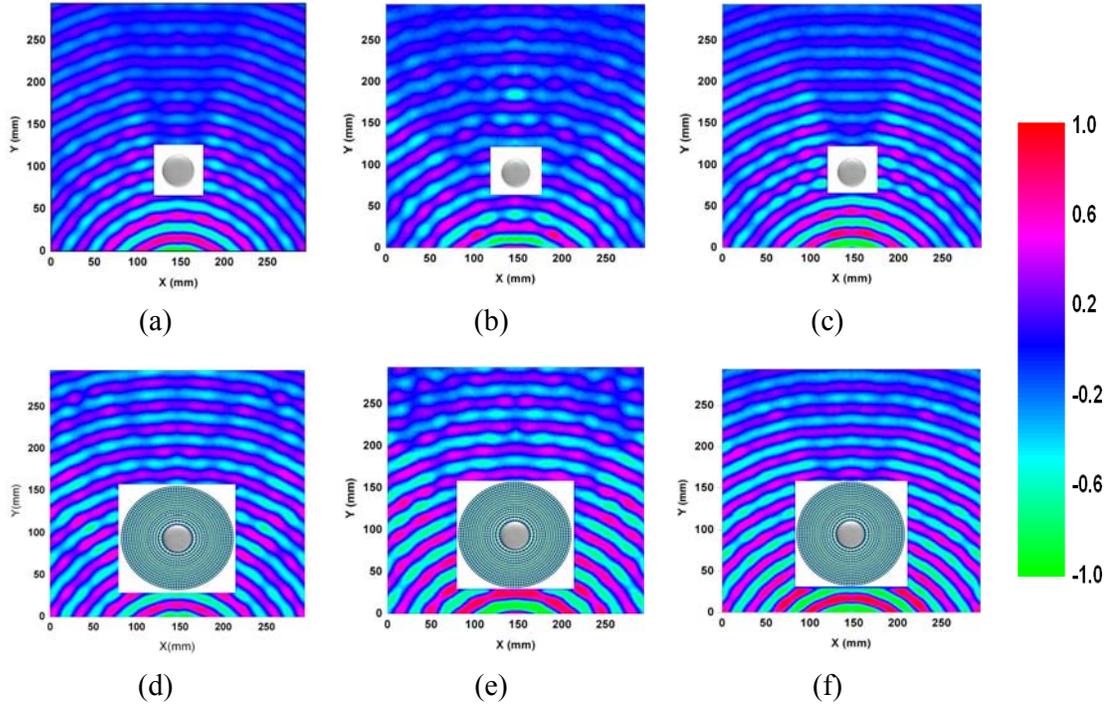

**Fig. 3.** Measured pressure field mappings of the bare steel cylinder and the cloaked steel cylinder illuminated with a point ultrasound source. The cloak lies in the center of the water tank and surrounds the steel cylinder. The scattering field patterns of the bare steel cylinder at (a) 60 kHz (b) 52 kHz and (c) 64 kHz. The pseudo colormaps in the immediate environment of the cloaked steel cylinder at (d) 60 kHz (e) 52 kHz and (f) 64 kHz.

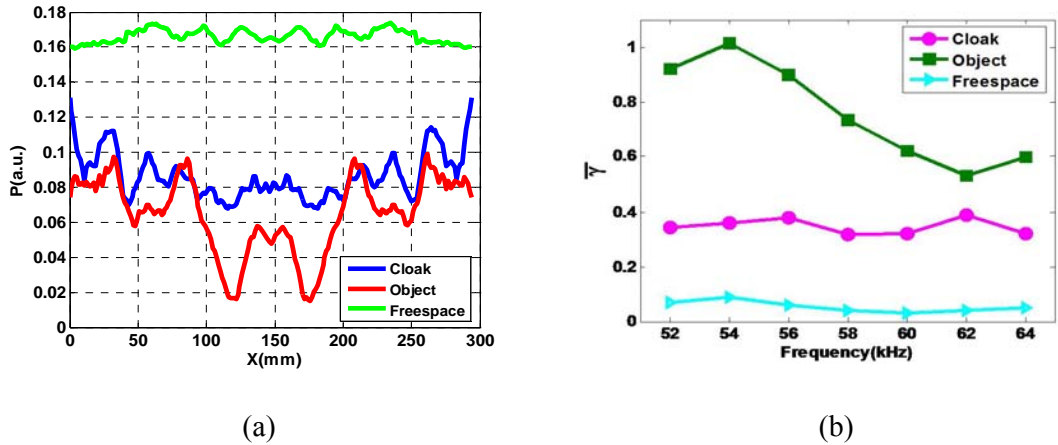

**Fig. 4.** Frequency dependence of the averaged visibility of the steel cylinder with and without the acoustic cloak. (a) The measured peak values of the pressure field along the wavefront lies between y=100 mm and y=170 with and without cloak at 60 kHz. The green line plot is the reference case when there is no object in the water tank. (b) Plot of the averaged visibility. The experimental results measured with and without cloak are marked by the magenta circles and green square respectively. The reference visibility when there is no object is marked by blue triangular.